\documentstyle[floats,aps,epsf,prl]{revtex}

\begin{document}
\draft
\twocolumn[\hsize\textwidth\columnwidth\hsize\csname @twocolumnfalse\endcsname

\title{Quasi-Particle density of states of disordered $d$-wave superconductors}

\author{Bodo Huckestein and Alexander Altland}

\address{Institut f\"ur theoretische Physik III,
Ruhr-Universit\"at Bochum, D-44780 Bochum, Germany}


\maketitle

\begin{abstract}
  We present a numerical study of the quasi-particle density of states
  (DoS) of two-dimensional $d$-wave superconductors in the presence of
  smooth disorder. We find power law scaling of the DoS with an
  exponent depending on the strength of the disorder and the
  superconducting order parameter in quantitative agreement with the
  theory of Nersesyan {\em et al.\/} (Phys.\ Rev.\ Lett. {\bf 72},
  2628 (1994)).  For strong disorder a transition to a constant DoS
  occurs. Our results are in contrast to the case of short-ranged
  disorder.
\end{abstract}
\pacs{PACS: 74.25.Bt, 74.20.Mn, 74.25.Jb, 72.15.Rn}
\vskip2pc

]

In recent years, motivated by its relevance to the physics of the
cuprates, $d$-wave superconductivity has become a subject of intensive
research. The key feature distinguishing $d$-wave superconductors from
their $s$-wave relatives is the existence of four zero energy 'nodes'
on the Fermi surface, in the vicinity of which low energy
quasi-particle excitations exist. This fermion system---four species
of relativistic quasi-particles, subject to weak static
disorder---has vital influence on  low energy transport and thermodynamic
properties and, therefore, must be a central element of any
comprehensive theory of the $d$-wave superconductor.

Irritatingly, it has proven excruciatingly difficult to reach
consensus on even basic characteristics of these states.  The extent
of disagreement is most clearly displayed in the debate on the energy
dependent mean quasi-particle density of states (DoS), $\rho(E)$: On
the one hand, application of self-consistent approximation
schemes\cite{gor'kov85,schmitt-rink86,hirschfeld86,lee93} or
non-perturbative approaches specific to certain realizations of the
disorder\cite{hettler96,balatsky96,pepin98}, has led to the
prediction of a finite or even diverging quasi-particle DoS at zero
energy. In contrast, field theory approaches to the
problem\cite{nersesyan94,senthil98}
categorically predict $\rho(E)=0$, as in the non-disordered Dirac
system.  Yet, even these theories within themselves come
to varying conclusions as to the energy dependence of the DoS for
$E\not=0$.

Recently it has become clear that much of this controversy roots in
the fact that, unlike with more conventional disordered fermion
systems, the standard paradigm of 'insensitivity of global observables
to microscopic details of the disorder' is apparently violated in the
$d$-wave system. Broadly speaking, two different categories of
disorder have to be distinguished: (i) hard scattering off $s$-wave
impurities, mixing the formerly isolated four low energy
quasi-particle sectors, and (ii) soft scattering which predominantly
leads to randomisation {\it within} these sectors. Which of these
categories is more relevant to the physics of the cuprates is not
straightforward to decide (see, however, our comments below), and both
have been investigated theoretically.  As for (i), there is now
overwhelming evidence that, apart from the case of asymptotically
strong impurities (impurities at the 'unitary
limit')\cite{lee93,balatsky96,pepin98,zhu}, the DoS vanishes as $E \to
0$\cite{nersesyan94,senthil98,atkinson}.  Away from zero energy a
variety of different DoS profiles, depending on the realization of the
disorder distribution, exist.

The subject of this Letter is a numerical study of the complementary
case, (ii). What makes this regime special, and why is it necessary to
discriminate between (i) and (ii) at all? The distinguishing feature
of the soft scattering regime is that the four low energy sectors are
decoupled. The absence of inter-node coupling has profound and
qualitative influence on the low energy properties of the system.
Indeed\cite{altland00.1}, it is the nodal coupling criterion, and not
so much the specifics of a short range correlated disorder
distribution\cite{atkinson}, that holds responsible for much
of the discrepancy between the field theory approaches to the problem.

Before turning to our numerical analysis of the quasi-particle
spectrum for soft scattering, let us briefly
summarize some key features of the system: Each node accommodates a
system of Dirac fermions subject to a random vector potential which
describes the stochastic low momentum transfer scattering. For low
energies and finite size systems -- the 'zero-dimensional' limit --
the properties of the system become fully universal and can be
described in terms of a suitably constructed random matrix theory
(RMT)\cite{verbaarschot93} (a). Due to the non-standard
symmetries of random gauge Dirac fermions (symmetries of class $A$III
in the terminology of Ref.~\cite{zirnbauer96}) this theory differs
profoundly from standard Wigner-Dyson RMT.  The opposite
extreme, thermodynamically extended systems is widely accessible to
analytical approaches, too\cite{mudry96,ludwig94}.
Specifically, for the $d$-wave problem Nersesyan, Tsvelik and Wenger
(NTW)~\cite{nersesyan94} have shown that the DoS scales as (b)
\begin{equation}
  \label{eq:6}
  \rho(E) \sim |E|^\alpha, \qquad \alpha =
  \frac{1-g}{1+g}, \qquad g=\frac{W^2}{16\pi\Delta t},
\end{equation}
where $W$ is the strength of the disorder, and $t$ and $\Delta$ are
the tight binding coupling strength and order parameter of the
superconductor, respectively. It has been argued\cite{gurarie} that
at $g=1$, i.e.  at the zero of the exponent in (\ref{eq:6}), a
transition to a qualitatively different phase 
takes place. On the strong disorder side of this transition, $g>1$, the DoS
is expected to be energy independent, $\rho(E) = {\rm const.}$ (c).
Further, (d), any amount of {\it hard} scattering coupling the low
energy nodes represents a marginally relevant
perturbation\cite{altland00.1} driving the system towards the coupled
regimes (i) mentioned above.

For completeness we mention that all these features find their common
origin in the fact that the low energy physics of isolated nodes is
described by a  Wess-Zumino-Witten (WZW) model on a group
manifold that depends on the treatment of the disorder
(replica\cite{nersesyan94,fendley} or
supersymmetry\cite{altland00.1}). In NTW's analysis of this
connection it has  erroneously been assumed
that the WZW action {\it globally} describes the low-energy
quasi-particle system, independent of the form factor of the
scattering. In fact, however, the WZW model is readily destabilized by
inter-node scattering which is one way of explaining the
aforementioned qualitative differences between cases (i) and (ii).
For a detailed account of the WZW-formulation, and its destruction, we
refer to Ref.\cite{altland00.1}.

Below we will put the phenomenology (a-d) to a numerical test.  Before
turning to a more detailed description of our analysis, let us
summarize the main results. We find that the large scale structure of
the energy dependent DoS can be characterized in terms of three
different regimes: For low energies $E$ above the chemical potential 
(owing to the Dirac structure of the problem, the DoS is
symmetric $\rho(E) = \rho(-E)$) the DoS profile is dominated by finite
size effects. At $E=0$ the DoS vanishes in a way which (for strong
enough disorder) is described by RMT.
The extent of the low energy
regime shrinks with increasing system size. For larger energies, it is
succeeded by a regime of power law scaling. Varying the two basic
parameters $\Delta/t$ and $W/t$ characterizing the
model, we find agreement with eq.~(\ref{eq:6}). For disorder strength
in excess of $g=1$, the DoS assumes a constant value, in accord with
the prediction of Ref.~\cite{gurarie}.  The scaling regime ends at
energies of the order of $\Delta$ where a non-universal high energy
regime, not considered in this Letter, begins. 
Upon lowering the correlation length of the disorder, the scaling
behaviour observed in the center portions of the band is rapidly
destructed.


We consider the lattice quasi-particle Hamiltonian
\begin{equation}
  \label{eq:1}
  H = \sum_{ ij;\sigma} (t_{ij}-\mu\delta_{ij})
  c_{i\sigma}^\dagger c_{j\sigma}^{\phantom{\dagger}}
  + \sum_{ ij} \Delta_{ij}
  c_{i\uparrow}^\dagger c_{j\downarrow}^\dagger + \mathrm{h.c.},
\end{equation}
with the hopping matrix elements $t_{ij}$, chemical potential $\mu$,
and order parameter $\Delta_{ij}$. The sums run over points of a
two-dimensional square lattice with spacing $a$ and the operators
$c_{i\sigma}^\dagger$ create a spin-1/2 particle of spin $\sigma$ at
site $i$. In the following, we take only into account on-site
potentials and nearest-neighbor hopping, $t_{ij}=\epsilon_i\delta_{ij}
+ t\delta_{i,j\pm e_k}$, where $e_k$ is the unit vector in
$k$-direction. For convenience we set $\mu=0$ (the half-filled band.)
The order parameter $\Delta_{ij}=\Delta(\delta_{i,j\pm
  e_x}-\delta_{i,j\pm e_y})$ has $d_{x^2-y^2}$-symmetry. 
In the following we will only consider disorder in the on-site
potentials $\epsilon_i$ and not determine the order parameter
self-consistently. (As pointed out in \cite{atkinson}, a
self-consistent determination of the order parameter may be necessary
to quantitatively compare with experimental data.) Finite
correlations in the disorder potential are introduced through
\begin{equation}
  \label{eq:4}
  \epsilon_i = \frac{W}{\sqrt{\Sigma}}\sum_j f_j
  \exp\left(-\frac{|{\mathbf{r}}_i-{\mathbf{r}}_j|^2}{\xi^2}\right),
\end{equation}
$\Sigma = \sum_j \exp(-2|{\mathbf{r}}_j|^2/\xi^2)$, where the $f_j$
are independent random variables, uniformly distributed in the
interval $[-1/2,1/2]$, and $W$ is a measure of the strength of the
disorder.  Taking $\xi=2a$ the coupling of the four Dirac nodes is
much weaker than the coupling of states in the vicinity of a single
node (by a factor of $\exp(-2\pi^2)$).

In the following, we take the lattice constant $a$ and the hopping
amplitude $t$ as the measures of length and energy, respectively. The
linear dimensions of the system vary between 15 and 45 and the
spectrum is averaged over 64 points in the first Brioullin zone. We
diagonalize the Hamiltonian (\ref{eq:1}) for 16 to 56 disorder
realizations and calculate the DoS $ \rho(E) = L^{-2} \sum_i
\delta_\Gamma(E - E_i)$, where $\delta_\Gamma(E)$ is a normalized
Gaussian of width $\Gamma$. A finite $\Gamma$ smoothes the DoS, but
also washes out narrow features.

\begin{figure}
  \begin{center}
    \epsfxsize=7.6cm
    \leavevmode
    \epsffile{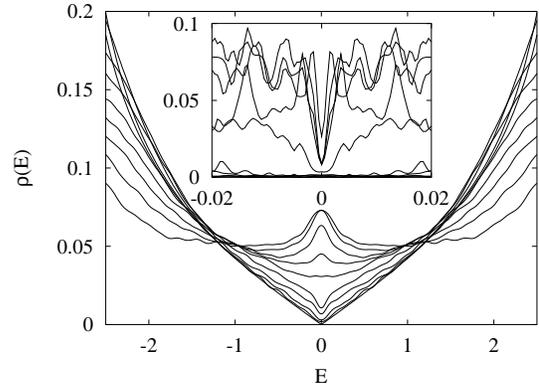}
    \caption{Density of states for disorder of strength $W=0,1,\dots,8,10$
      (bottom to top at $E=0$), correlation length $\xi=2$, and
      $\Delta=1$. The system size is $L=33$ and the broadening
      $\Gamma=0.05$. The inset shows the same data on a smaller scale
      with $\Gamma=0.0005$. The finite DoS at $E=0$ is due to the
      finite broadening $\Gamma$.}
    \label{fig:1}
  \end{center}
  \vspace{-0.5cm}
\end{figure}
Fig.~\ref{fig:1} shows the DoS for various values of disorder. The
three regimes mentioned earlier are the low-energy region
where the bump develops, the intermediate regime up to the approximate 
crossing point near $E=1$, and the high-energy regime beyond. We will
first discuss the most interesting intermediate regime before turning
our attention to low energies.
\begin{figure}
  \begin{center}
    \epsfxsize=7.6cm
    \leavevmode
    \epsffile{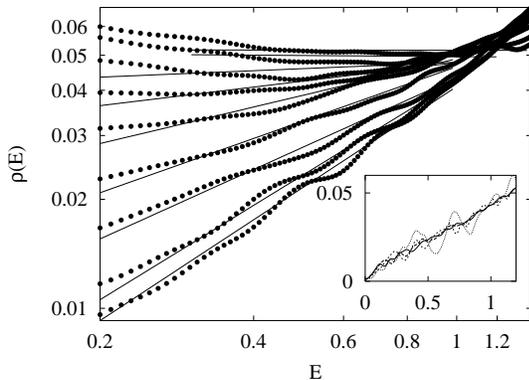}
    \caption{Double logarithmic plot of the density of states of
      Fig.~\ref{fig:1}. Disorder ranges from $W=1$ to 10. Dots
      ($\bullet$) represent data and lines power law fits to the
      respective intervals. Inset: Density of states for $W=2$ and
      $L=15$ (dotted), 25 (short-dashed), 35 (long-dashed), and 45
      (solid). Note that the numerical uncertainties are considerably
      smaller than the amplitude of the fluctuations.}
    \label{fig:2}
  \end{center}
  \vspace{-0.5cm}
\end{figure}
In order to compare our data to the results of NTW, we fit the data to
power laws $\rho(E)\propto E^\alpha$ in a interval $[E_{\rm
  min},E_{\rm max}]$. $E_{\rm min}$ is choosen such as to exclude the
first maximum. From the inset in Fig.~\ref{fig:2} it is clear that
this feature as well as the fluctuations are finite-size effects that
appear to vanish in the large system limit. The upper limit $E_{\rm
  max}$ is a high energy cut-off of the order of $\Delta$ beyond which
the NTW theory  no longer applies. At the systems sizes
considered in this work, the ratio $E_{\text{max}}/E_{\text{min}}$ is
about 5.  This is a rather narrow range to establish a power law.
Nevertheless, we feel that our procedure is justified in the present
case, as it is not just a single power law with a single exponent that
we are dealing with. Instead, eq.~(\ref{eq:6}) predicts that there is
a whole family of power laws with the exponents depending in a unique
way on the parameters $W$ and $\Delta$. It is the agreement of this
whole functional dependence that gives us confidence in the validity
of our analysis even when the establishment of every single power law
might be questionable.
\begin{figure}[tbp]
  \begin{center}
    \epsfxsize=7.6cm
    \leavevmode
    \epsffile{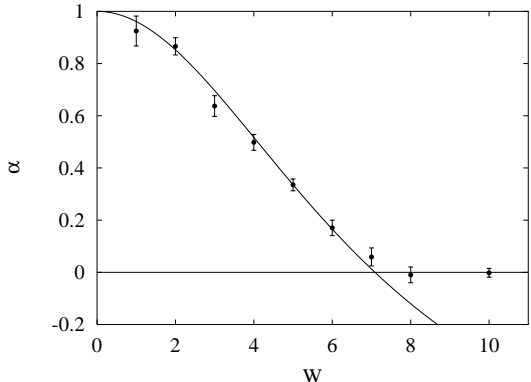}
    \caption{Exponents $\alpha$ extracted from the fitted curves in
      Fig.~\ref{fig:2} as a function of disorder $W$ for
      $\Delta=1$. The solid curve is the result of NTW,
      eq.~(\ref{eq:6}).}
    \label{fig:4}
  \end{center}
  \vspace{-0.5cm}
\end{figure}
Figure~\ref{fig:4} shows the exponents $\alpha$ extracted from the
fitted curves in Fig.~\ref{fig:2} together with the result of
eq.~(\ref{eq:6}) of \cite{nersesyan94}.  A good agreement is apparent
up to a disorder strength of $W\approx 7$ where the NTW exponent
changes sign. Numerically, we do not find a divergent DoS at stronger
disorder but rather a finite value ($\alpha=0$) as predicted by
Gurarie \cite{gurarie}.

\begin{figure}[tbp]
  \begin{center}
    \epsfxsize=7.6cm
    \leavevmode
    \epsffile{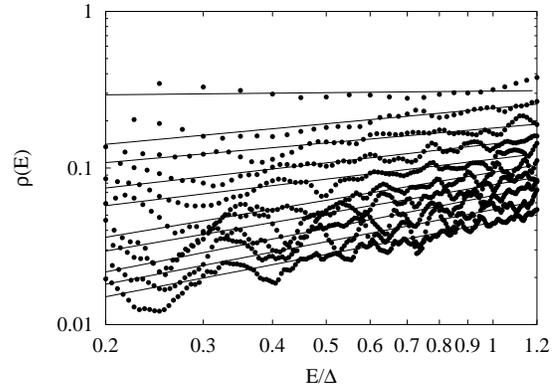}
    \caption{Density of states for order parameter $\Delta=0.1$ to 1.0
      (top to bottom) and disorder $W=3$. Each curve is shifted by a
      factor of 1.2 for clarity.}
    \label{fig:7}
  \end{center}
  \vspace{-0.5cm}
\end{figure}

\begin{figure}[tbp]
  \begin{center}
    \epsfxsize=7.6cm
    \leavevmode
    \epsffile{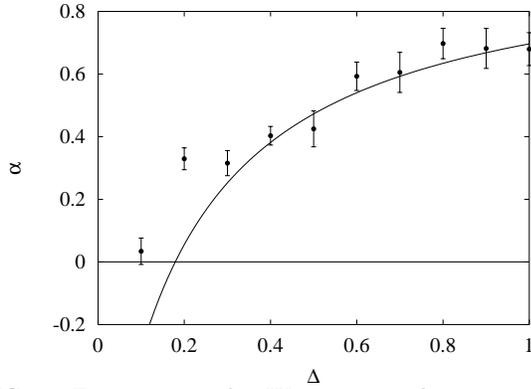}
    \caption{Exponents $\alpha$ for $W=3$ as a function of order
      parameter $\Delta$. The solid curve is the result of NTW,
      eq.~(\ref{eq:6}).}
    \label{fig:8}
  \end{center}
  \vspace{-0.5cm}
\end{figure}

To further test eq.~(\ref{eq:6}) we fix the disorder strength $W=3$
and vary the order parameter $\Delta$. Figure~\ref{fig:7} shows the
DoS for $\Delta$ between 0.1 and 1.0 as a function of $E/\Delta$. This
rescaling takes care of the fact that the mean level spacing of the
clean system is proportional to $\Delta$. The anisotropy dependence of 
the fitted exponents is shown in Fig.~\ref{fig:8}. Again, reasonably
good agreement with eq.~(\ref{eq:6}) is found for $\Delta>0.2$ while a 
constant and not a diverging DoS is found at $\Delta=0.1$. 

The inset of Fig.~\ref{fig:1} shows that the DoS does indeed vanish at
$E=0$. At weak disorder ($W\leq4$) we see a remnant of the clean
spectrum. Here the disorder is too weak to couple neighboring states
in momentum space. At $W\approx5$ this coupling exceeds the level
separation and the universal RMT behaviour for systems of class
$A$III, with a DoS 'microgap' linear in energy, develops\cite{fn1}. 

To conclude the central part of our analysis, we have presented a
numerical study of the quasi-particle spectrum in $d$-wave
superconductors with soft scattering. In the regime of moderate
disorder strength, we obtain scaling of the DoS that agrees
quantitatively with the analytical results of NTW.
For strong  disorder, the analysis confirms the recent
prediction\cite{gurarie} of an energetically constant background DoS.
In a way our analysis is complementary to recent studies of hard
scattering systems\cite{atkinson} and we are left with the question
which of these alternatives is more relevant to the physics of 'real'
cuprates. It is probably difficult to give a universally applicable
answer to this question. On the face of it, disorder in high $T_c$
superconductors is due to small metallic donors, e.g.  Zn-impurities,
in favour of the hard variant, (i). (Concrete evidence for the
presence of hard scattering in Zn-doped systems is provided by the
experimental observation\cite{hudson99} of so-called mid-gap
resonances [i.e.  resonances due to bound states forming in the
immediate vicinity of a strong local impurity]).  On the other hand,
the very existence of a $d$-wave phase, stabilized through a mechanism
unknown at present, would not be compatible with too strong an amount
of---pairbreaking---hard scattering, i.e. the renormalized {\it
  effective} potential seen by the quasi-particle states may well
carry characteristics of type (ii) and be predominantly soft (see
Refs.\cite{kulic97} for a more elaborate discussion
of this point.) Equally important, the net features of the
quasi-particle system are not only determined by the fixed microscopic
structure of the disorder background but also depend on temperature
and observation energy. E.g.  consider a system with a certain
residual amount of hard scattering superimposed on a predominantly
soft background. For quasi-particle energies in excess of the the
inter-node scattering rate, the coupling between the nodes is
inessential and the characteristics of the soft system will prevail.
Lowering the energies, a crossover towards the hard system takes
place. 

Although the picture above suggests, that 'real life' systems will
typically display complex crossover behaviour, recent progress in
purely {\it theoretical} understanding of disorder in $d$-wave
superconductors has been tremendous. It seems to be clear now, that
much of the controversy that developed around the profile of the
quasi-particle DoS is related to non-congruent modellings of the
disorder. Indeed, the majority of theoretical approaches to the
problem sits comfortably with one of the disorder realizations
investigated numerically in this work or in complementary
papers\cite{zhu,atkinson}. (There is one prominent exception to that
rule, viz. Refs.\cite{hettler96} where a finite DoS for continuously
distributed disorder was predicted. To our understanding the
discrepancy is explained by the peculiar lattice implementation
underlying these papers (see Ref.\cite{altland00.1} for a more
elaborate discussion of this point) which implies that no
superconductor is modelled.)  Broadly speaking, there seem to be three
categories of disorder that have to be distinguished: binary alloy
type scatterer at the unitary limit, large momentum transfer
scatterers of non-unitary type and soft disorder. In spite of the
relative---as compared to normal metals---complexity of this
classification, the categories within themselves still display a large
amount of universality.

{\it Acknowledgement:} We than B.D. Simons for discussions and
M. Janssen for reading the manuscript.


\begin{references}
\bibitem{fn1}
That the low-energy regime is {\it not} described by the ntw formalism follows
  from the fact that the latter has been developed for infinte systems. more
  precisely, the ntw theory is characterized by some correlation length
  $\xi(e)$ which increases upon lowering $e$. as soon as $\xi(e)\sim$ (system
  size), a crossover to a low energy regime dominated by finite size effects
  takes place.
\end{references}

\end{document}